\documentclass[acmtog,nonacm]{acmart}

\usepackage{times}
\usepackage{epsfig}
\usepackage{graphicx}
\usepackage{amsmath}

\usepackage{multirow}
\usepackage{subfigure}
\usepackage{wrapfig}
\usepackage{booktabs}  

\begin{document}

\title{SonicRadiation: A Hybrid Numerical Solution for Sound Radiation without Ghost Cells}
\author{Xutong Jin}
\email{jinxutong@pku.edu.cn}

\affiliation{%
  \institution{School of Computer Science, Peking University}
 \country{China}
}

\author{Fei Zhu}
\authornote{corresponding author}
\email{feizhu@pku.edu.cn}
\affiliation{%
  \institution{School of Computer Science, Peking University}
 \country{China}
}

\author{Guoping Wang}
 \email{wgp@pku.edu.cn}
\affiliation{%
  \institution{School of Computer Science, Peking University}
  \country{China}
}

\author{Sheng Li}
\authornote{corresponding author}
\email{lisheng@pku.edu.cn}
\affiliation{%
  \institution{School of Computer Science, Peking University}
  \country{China}
}

\begin{abstract}
Sound radiation simulation, a key component of physically based sound synthesis, has posed challenges in the context of complex object boundaries. Previous methods, such as ghost cell-based finite-difference time-domain (FDTD) wave solver, have struggled to address these challenges, leading to large errors and failures in complex boundaries because of the limitation of ghost cells. 
We present SonicRadiation, a hybrid numerical solution capable of handling complex and dynamic object boundaries in sound radiation simulation without relying on ghost cells. We derive a consistent formulation to connect the physical quantities on grid cells in FDTD with the boundary elements in the time-domain boundary element method (TDBEM). Hereby, we propose a boundary grid synchronization strategy to seamlessly integrate TDBEM with FDTD while maintaining high numerical accuracy. Our method holds both advantages from the accuracy of TDBEM for the near-field and the efficiency of FDTD for the far-field. Experimental results demonstrate the superiority of our method in sound radiation simulation over previous approaches in terms of accuracy and efficiency, particularly in complex scenes, further validating its effectiveness.
\end{abstract}


\begin{teaserfigure}
\centering
\includegraphics[width=0.8\linewidth]{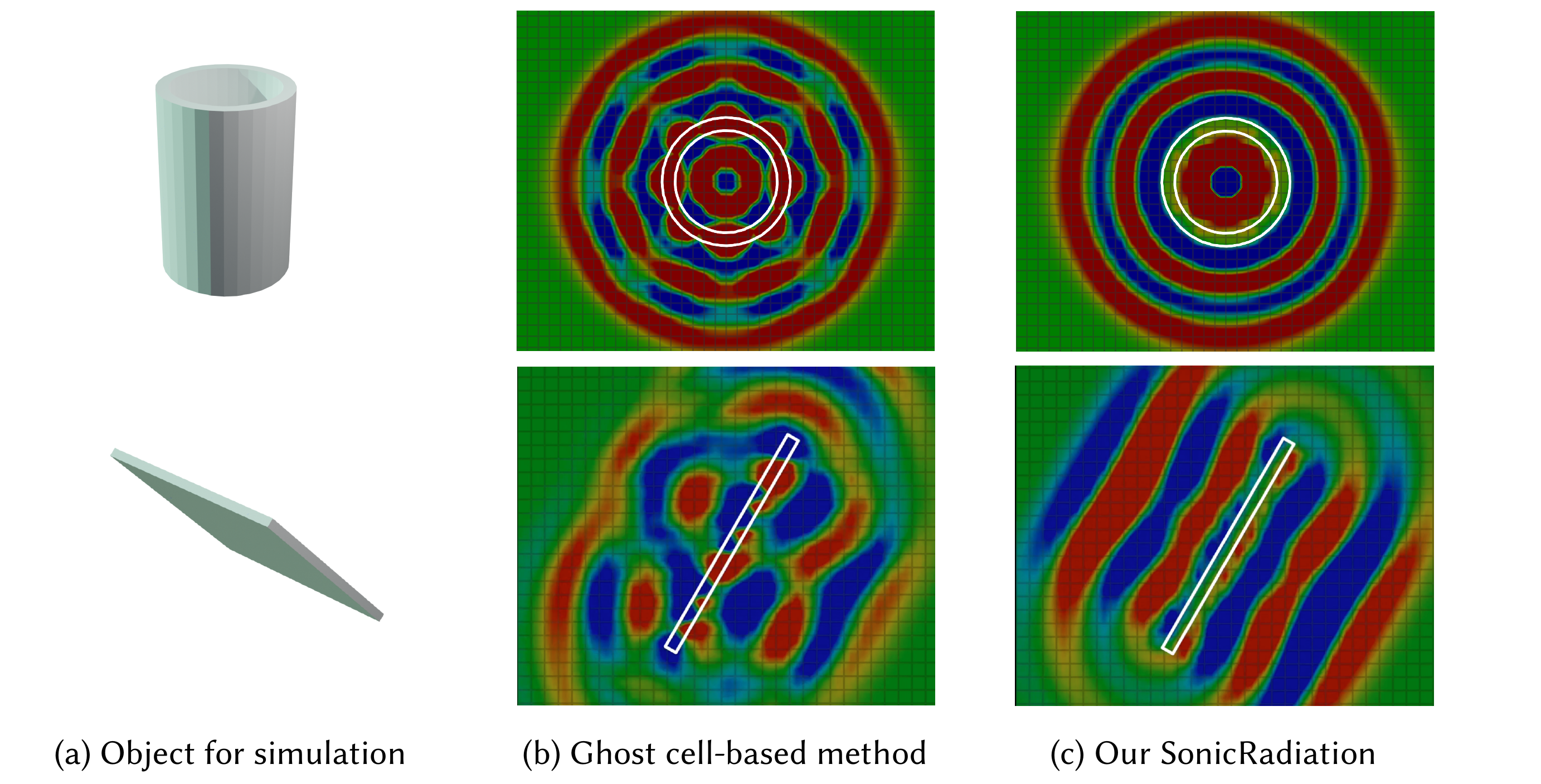}    
\caption{
Sound radiation comparison between the ghost cell–based method (second-order boundary approximation)~\cite{Wang2018} and our \textit{SonicRadiation} under the same grid resolution.
(\textbf{a}) Test configurations: a vibrating tube and a vibrating plane at 4000 Hz with uniform amplitude excitation, which theoretically should generate a spatially uniform pressure field. In the visualizations, red, blue, and green denote positive, negative, and zero pressure values, respectively; white outlines the object boundary; and gray lines indicate the FDTD grid. Under identical simulation settings, the ghost cell–based method (\textbf{b}) exhibits noticeable inaccuracies, whereas our approach (\textbf{c}) produces results that are both more accurate and computationally more efficient. Please refer to the accompanying video for more details.
}
\label{fig:teaser}
\end{teaserfigure}

\maketitle






\section{Introduction}
Physically based sound synthesis is a technique used to create the sound of virtual sound sources. A central aspect of sound synthesis involves the accurate simulation of sound radiation, the process by which sound waves are emitted from a source and propagate through the environment. Sound radiation significantly contributes to the overall quality and realism of synthesized sounds and has, therefore, been the focus in the field of physically-based sound synthesis. 

Frequency-domain boundary element methods have been widely used in previous research to calculate sound radiation~\cite{Kirkup2019, PAT, Li2016}. While this approach has proven effective, it lacks the ability to handle time-domain information in dynamic scenes. Finite-difference time-domain (FDTD) solver is an alternative for sound radiation in dynamic scenes~\cite{larsson2003,botteldooren1995}. However, these methods struggle to handle complex or dynamic boundaries accurately. Wang~\cite{Wang2018} proposed a ghost cell-based FDTD method that can handle more complex boundary conditions. However, it requires a high grid resolution for rasterizing objects and locating solid cells with air cell neighbors as ghost cells. Additionally, it needs to ensure the mirror point of the ghost cell relative to the boundary in the air. As a result, the ghost cell-based FDTD method is slow and can introduce geometric problems. Although the time-domain boundary element method (TDBEM)~\cite{langer2008,Banjai2008,Banjai2010} is able to accurately calculate the sound radiation of dynamic scenes, its complexity increases significantly for long-range propagation and large numbers of boundary elements. Therefore, it typically requires extensive computing resources with costly time overhead.

Our method is dedicated to addressing sound radiation challenges in complex scenarios, such as dynamic objects, thin shells, and intricate boundaries. We aim to harness the power of time-domain boundary element methods (TDBEM) in handling complex boundaries instead of the ghost cell-based method while benefiting from the computational efficiency of finite difference time domain (FDTD) methods that use large grid cell sizes.

In this paper, we introduce SonicRadiation, a novel hybrid solver that combines the time-domain boundary element method (TDBEM) with finite-difference time-domain (FDTD) solver for physically accurate sound radiation simulation. We establish a framework that unifies the treatment of Dirichlet and Neumann data of boundary elements and pressure values in grid cells. This approach leverages the strengths of TDBEM for near-field interactions and FDTD for far-field contributions. We present a boundary-grid synchronization strategy that efficiently computes far-field values for grid cells using the Dirichlet and Neumann data of nearby boundary elements, as well as interpolates far-field values for updating the Dirichlet data of all boundary elements. Our method is adept at handling dynamic boundaries seamlessly without additional complexity. Compared with the ghost cell-based method, our approach is faster, eliminates the need for rasterization of objects, and circumvents the geometric issues associated with ghost cells.

We provide analytical test cases to validate our approach, including monopole tests and acoustic transfer tests, under various experimental conditions, as highlighted in \autoref{fig:teaser}. Our numerical solver excels at synthesizing reliable sound-radiation effects with exceptional efficiency. Furthermore, our hybrid approach holds promise for enhancing numerical solutions in grid-based physical simulations.

\section{Related work}
The use of physically based methods for sound simulation has a long history in computer sound and music~\cite{smith1992, Gaver1993, cook2002}. This includes the synthesis, propagation, and rendering of audio effects~\cite{DineshSurvey}. Our focus is primarily on the simulation of near-field sound propagation, which must be integrated with sound synthesis techniques to achieve the desired result.

Modal sound synthesis is a widely used technique for synthesizing the sounds of rigid bodies~\cite{van2001, OBrien2002, Raghuvanshi2006}. More advanced modal sound synthesis models have also been proposed, such as acceleration noise synthesis models~\cite{Chadwick2012}, contact models~\cite{contact}, and data-driven models~\cite{Ren2013, DeepModal, NeuralSound}. In addition, physically based sound synthesis has been applied to other types of objects such as thin shells~\cite{HarmonicShell} and fluids such as water~\cite{Minnaert1933, Moss2010, HarmonicFluids}.

The sound propagation that occurs after sound synthesis takes into account the effects of sound reflection and diffraction caused by the object itself and other nearby objects. A commonly used method to solve the field of sound radiation from a linear sound model from the Helmholtz equation is the boundary element method (BEM) of the frequency domain ~\cite{Kirkup2019}. Precomputed acoustic transfer (PAT)~\cite{PAT} is another method that uses precomputed radiation solutions from BEM to quickly estimate acoustic pressure amplitudes. Other techniques that have been proposed include the use of a single-point multipole expansion to represent the sound source with fewer coefficients~\cite{Langlois2014, Zheng2010, Rungta2016}, or Far-Field Acoustic Transfer (FFAT) maps~\cite{HarmonicShell, KleinPAT}, which can speed up the rendering process. However, these Helmholtz equation-based methods have limitations, such as the assumption of linear modal dynamics (independent mode) and the lack of ability to acoustically interact with nearby objects. Our method does not have these limitations.

Finite-difference time-domain (FDTD) methods~\cite{larsson2003} have been used to approximate solutions to the wave equation and have been used to solve broadband acoustic radiation problems in various applications such as computational room acoustics~\cite{botteldooren1995,Bilbao2013} and musical instrument design~\cite{bilbao2009}. They have also been used to achieve high throughput~\cite{MEHRA201283, Micikevicius2009, Allen2015}. However, medium interfaces can introduce errors, and Wang~\cite{Wang2018} proposed a second-order approximation ghost cell method to achieve higher accuracy. However, their method still faces difficulties in handling complex interfaces due to the necessity of object rasterization. our method can handle any object interface without the need for rasterization while maintaining efficiency and accuracy.

In sound propagation, the motion of sound sources and surfaces is crucial. The time-domain boundary element method (TDBEM)~\cite{langer2008} is an extension of the frequency domain BEM that is able to handle such problems. Convolution quadrature method (CQM)~\cite{Lubich1994, Banjai2010, Banjai2008} has gained popularity due to its unconditional stability and ease of computation. However, it has high computational complexity and TDBEM is too slow for complex scenes. The Fast Multipole Method (FMM)~\cite{liu2009} can reduce the computational cost of BEM, but it remains slow due to difficulties in parallelization and high implementation complexity. Vortex-in-cell (VIC) methods~\cite{COUET1981} and particle-particle-mesh (PPPM) methods~\cite{ANDERSON1986,Zhang2014} can speed up fluid dynamics by approximating smooth long-range interactions using finite differences. Similarly, our method approximates the smooth long-range interactions in TDBEM with FDTD.

\section{Preliminary}
\label{sec:background}
The finite difference time domain (FDTD) solver and the time-domain boundary element method (TDBEM) are both instrumental in solving the sound pressure distribution based on dynamic boundary conditions. In this section, we will provide an introduction to these two methods, as they play a pivotal role in our algorithm.

\subsection{Finite Difference Simulation}
\label{FDTD}
The finite difference time domain method originates from the acoustic wave equation, which describes the evolution of pressure perturbations in an open set containing an acoustic medium:
\begin{equation}
\frac{\partial^2 p(\mathbf{x},t)}{\partial t^2} = c^2 \nabla^2 p(\mathbf{x},t), \quad \mathbf{x} \in \Omega
\label{eq:wave_equation}
\end{equation}
Here, $p(\mathbf{x},t)$ represents the pressure at position $\mathbf{x}$ and time $t$, $\Omega$ denotes the open set containing the acoustic medium, and $c$ represents the velocity of sound in the medium.

The finite difference time domain method involves discretizing both space and time into a regular grid, and utilizing finite differences to approximate the spatial and temporal derivatives. For a grid with cell size $h$ and time step size $\tau$, the pressure update of grid cell $(u,v,w)$ at time step $n$, is given by:
\begin{equation}
p^{n+1}_{u,v,w} = (2 + c^2 \tau^2 \tilde{\nabla}^2) p^n_{u,v,w} - p^{n-1}_{u,v,w}\ ,
\label{eq:fdtd}
\end{equation}
where the discrete Laplacian, $\tilde{\nabla}^2$, is defined as:
\begin{align}
h^2\tilde{\nabla}^2 p_{u,v,w} &= p_{u-1,v,w} + p_{u+1,v,w} + p_{u,v-1,w} + p_{u,v+1,w}\nonumber\\ & + p_{u,v,w-1}+ p_{u,v,w+1} - 6p_{u,v,w}\ .
\label{eq:fdtd_laplacian}
\end{align}
By recursively solving \autoref{eq:fdtd}, the sound pressure of all grid cells can be determined. However, the algorithm complexity is $O(N^3)$ for a grid resolution of $N^3$. To maintain a reasonable time complexity, the grid resolution cannot be too high.

\subsection{Time-Domain Boundary Element Method}
\label{TDBEM}
The time-domain boundary element method (TDBEM) is an extension of the traditional frequency domain boundary element method (BEM). In this paper, we specifically focus on the convolution quadrature method (CQM)~\cite{Lubich1994} based TDBEM, with its detailed derivation available in~\cite{Banjai2008, Banjai2010}.

Let's assume that all object surfaces are discretized into $M$ triangular boundary elements, and consider a discretized time step size of $\tau$. Denote $\mathbf{\Phi}_j$ and $\mathbf{G}_j$ as vectors of length $M$ representing the discretized Dirichlet and Neumann data of all boundary elements at time step $j$. In our scenario, the discretized Neumann data $\mathbf{G}_j$ is known (obtained from the surface vibration acceleration), while the discretized Dirichlet data $\mathbf{\Phi}_j$ is unknown but can be determined from the discretized boundary integral equation:
\begin{equation}
    \left(\frac{1}{2} \mathbf{I} - \mathbf{D}_{0}\right) \mathbf{\Phi}_n = \sum_{j=1}^{L} \mathbf{D}_{j}\mathbf{\Phi}_{n-j} -\sum_{j=0}^L \mathbf{V}_{j} \mathbf{G}_{n-j}\ ,
\label{eq:discretized_boundary_equation}
\end{equation}
where $\mathbf{I}$ is a diagonal matrix with elements representing the area of each boundary element. Additionally, $\mathbf{V}_j$ and $\mathbf{D}_j$ are dense matrices of size $M \times M$ derived from the single layer boundary matrices and double layer boundary matrices in classical BEM~\cite{bempp}. Here, $L$ denotes the history time step.

By recursively solving \autoref{eq:discretized_boundary_equation}, all historical Dirichlet data $\mathbf{\Phi}_j$ can be determined. Then, the sound pressure at position $x$ and time step $n$ is computed as follows: 
\begin{equation}
    p_x^n = \sum_{j=0}^L  \left( \mathbf{d}_j(x) \cdot \mathbf{\Phi}_{n-j}-\mathbf{v}_j(x) \cdot \mathbf{G}_{n-j} \right)  \ ,
    \label{eq:discretized_potential_equation}
\end{equation}
where $\mathbf{v}_j(x)$ and $\mathbf{d}_j(x)$ are vectors of size $m$. 

When solving TDBEM, the most significant time consumption arises from \autoref{eq:discretized_boundary_equation}, primarily due to the dense matrices $\mathbf{V}_j$ and $\mathbf{D}_j$ with the size of $M \times M$. Consequently, the overall complexity is $O(LM^2)$. In many cases, the time complexity of TDBEM, $O(LM^2)$, can be much greater than that of FDTD, which operates at $O(N^3)$.



\begin{figure*}[ht]
\centering
\includegraphics[width=0.9\linewidth]{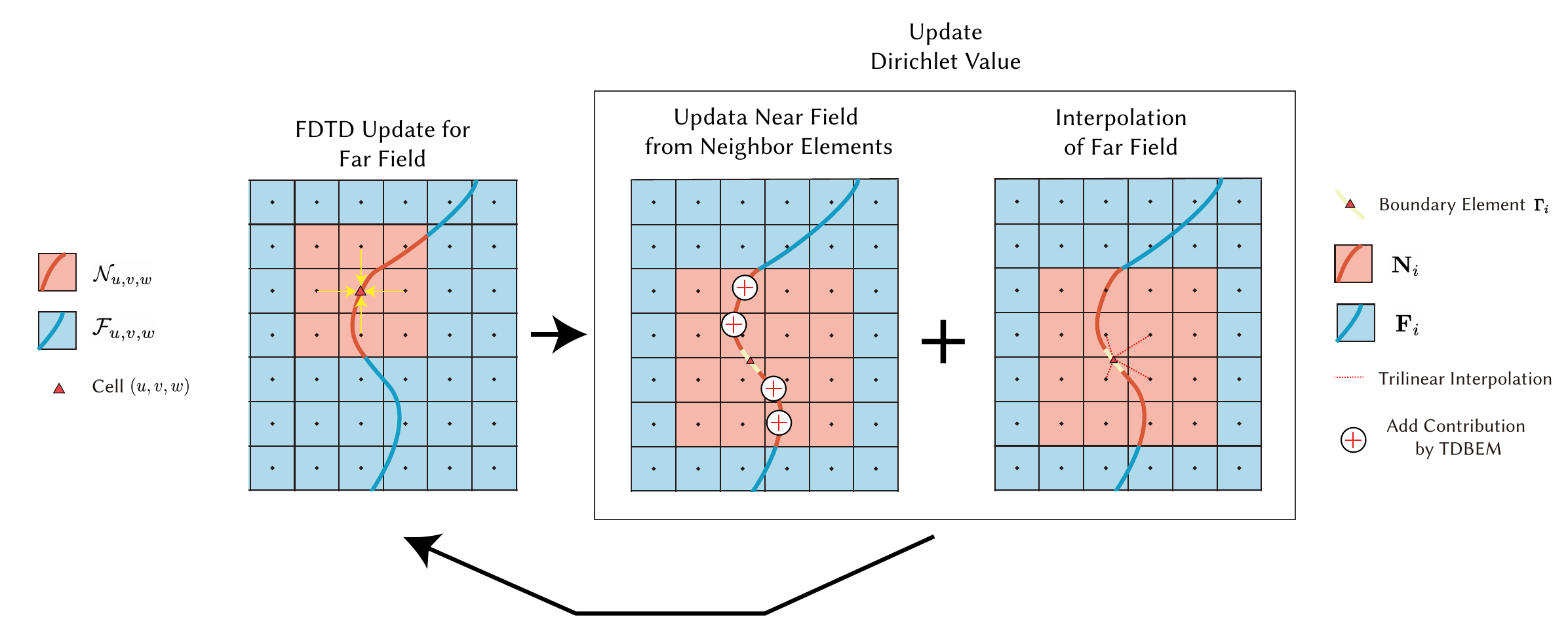}
\caption{
Overview of one step in the loop iteration of our approach: first, we utilize FDTD to update the far-field value for each grid cell. The far-field value for cell $(u, v, w)$ representing the summation of contribution from all boundary elements in $\mathcal{F}_{u,v,w}$. Next, for updating the Dirichlet values of boundary element $\Gamma_i$, we calculate the summation of contributions from boundary elements in $\mathbf{N}_i$ using TDBEM, and determine the summation of contribution from boundary elements in $\mathbf{F}_i$ through trilinear interpolation of the far-field values at surrounding grid cells.
}
\label{fig:overview}
\end{figure*}

\section{Our Method}
\label{sec:BEM-FDTD}
For the dynamic sound radiation problem, previous approaches generally employ FDTD to solve the sound pressure, utilizing the ghost cell method to approximate the boundary conditions ~\cite{Wang2018}. To maintain a reasonable time complexity, the grid resolution should be kept as low as possible.
However, a low-resolution grid can lead to noticeable errors when approximating the boundary conditions using the ghost cell method (\autoref{fig:teaser} (b)). our approach addresses this issue by a combined FDTD solver with TDBEM, which allows for more accurate boundary condition approximations and faster computations compared to the ghost cell-based method (\autoref{fig:teaser} (c)).


Assuming all historical data are known, our SonicRadiation needs to update all values in grid cells and all Dirichlet values of all boundary elements within an iteration following two steps:
\begin{enumerate}
    \item We implement a FDTD iteration without incorporating ghost cells (Sec.~\ref{subsec:fdtd_update}). The values of each grid cell are designated as the far-field value, considering only the contributions from boundary elements that are far from the grid cell itself. It is imperative to have access to the historical Neumann and Dirichlet values of boundary elements for the computation of the FDTD iteration.

    \item We employ the interpolation of the far-field values from all grid cells to approximate the contributions from boundary elements that are distant from a given boundary element. This allows for the efficient update of Dirichlet values for all boundary elements (Sec.~\ref{subsec:interpolate_far_field}).
\end{enumerate}
The comprehensive overview is depicted in \autoref{fig:overview}. The subsequent subsections will provide a detailed description of each of these steps.

\subsection{FDTD Update for Far-field Value}
\label{subsec:fdtd_update}
First, we apply a FDTD iteration on all grid cells. In this step, all grid cells are regarded as air cells. Therefore, we eliminate the need for ghost cells and their limitations. 

To explain how FDTD results can remain correct without relying on solid and ghost cells, we will start by examining some key observations. We can reinterpret the value of each grid cell as the summation of contributions from individual boundary elements by rewriting \autoref{eq:discretized_potential_equation} as:
\begin{equation}
p_x^n=\sum_{i=0}^M\sum_{j=0}^L\left([\mathbf{d}_j(x)
]_i[\mathbf{\Phi}_{n-j}]_i-[\mathbf{v}_j(x)]_i [\mathbf{G}_{n-j}]_i  \right) \ ,
\label{eq:FDTD_potential}
\end{equation}
where $[\cdot]_i$ represent the $i$-th element in a vector and $x$ is the center of the grid cell $(u,v,w)$. We denote $p_{u,v,w}^n(\Gamma_i)$ as the contribution of $i$-th boundary element $\Gamma_i$ to the pressure value of grid cell $(u,v,w)$ at time step $n$:
\begin{equation}
    p_{u, v, w}^n (\Gamma_i) = \sum_{j=0}^L\left([\mathbf{d}_j(x)
]_i[\mathbf{\Phi}_{n-j}]_i-[\mathbf{v}_j(x)]_i [\mathbf{G}_{n-j}]_i  \right) \ .
    \label{eq:FDTD_potential_single}
\end{equation}
In this paper, we define a boundary element in a cell as one where the center point of the boundary element lies within the grid cell. For the grid cell $(u,v,w)$, we denote the set of boundary elements in its $R_1 \times R_1 \times R_1$ neighbor cells as $\mathcal{N}_{u,v,w}$, and the set of other boundary elements as $\mathcal{F}_{u,v,w}$. For any set of boundary elements $\mathcal{S}$ (e.g., $\mathcal{N}_{u,v,w}$ or $\mathcal{F}_{u,v,w}$), we denote the summation of contributions from each boundary element in $\mathcal{S}$ to grid cell $(u, v, w)$ as:
\begin{align} 
    p_{u,v,w}^n(\mathcal{S}) = \sum_{\Gamma_i \in \mathcal{S}} p_{u,v,w}^n(\Gamma_i)\ ,
    \label{eq:compute_U}
\end{align}
Under this perspective, the full sound pressure value at the center of any grid $(u,v,w)$ can be represented as the summation of near-field part and far-field part:
\begin{equation}
    p_{u,v,w}^n = p_{u,v,w}^n( \mathcal{N}_{u,v,w}) +  p_{u,v,w}^n( \mathcal{F}_{u,v,w})\ .
\end{equation}
The core motivation to get rid of ghost cells in FDTD is that we only compute the far-field value $p_{u,v,w}^n( \mathcal{F}_{u,v,w})$ instead of the full value $p_{u,v,w}^n$ in all grid cells. 

As the Green function and its derivatives used in the boundary integration equation are all basic solutions to the Helmholtz equation, the sound pressure field which is only contributed from one boundary element or any set of boundary elements can also be approximated by the FDTD simulation. 

\begin{figure*}[ht]
    \centering
    \includegraphics[width=0.8\linewidth]{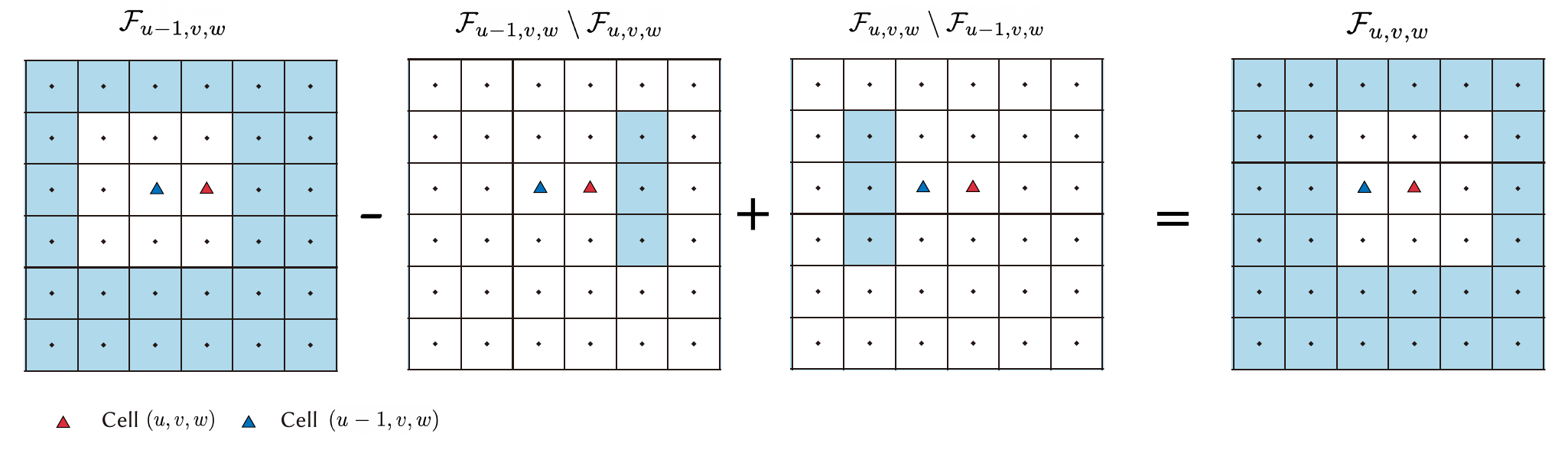}
    \caption{Illustration of the correction from far-field value defined in one cell to another neighboring cell. As the set of related boundary elements in the correction item (\(\mathcal{F}_{u-1,v,w} \setminus \mathcal{F}_{u,v,w}\) and \(\mathcal{F}_{u,v,w} \setminus \mathcal{F}_{u-1,v,w}\)) is small, the correction operation is fast.}
    \label{fig:fdtd_correction}
\end{figure*}

During the computation of the far-field value of cell $(u,v,w)$, we only consider the contribution from boundary elements in set \( \mathcal{F}_{u,v,w} \). Then, a FDTD iteration can be used to update the far-field value \( p_{u,v,w}^n( \mathcal{F}_{u,v,w}) \):
\begin{align}
    p_{u,v,w}^n( \mathcal{F}_{u,v,w}) &=  (2p_{u,v,w}^n( \mathcal{F}_{u,v,w}) -  p_{u,v,w}^{n-1}( \mathcal{F}_{u,v,w})) \nonumber \\ 
    &+ c^2 \tau^2 (\tilde{\nabla}^2p^n_{u,v,w}( \mathcal{F}_{u,v,w})) \ .
    \label{eq:FDTD_far_update}
\end{align}
In the right-hand side of this equation, the first part \( (2p_{u,v,w}^n( \mathcal{F}_{u,v,w}) -  p_{u,v,w}^{n-1}( \mathcal{F}_{u,v,w})) \) can be directly obtained from the grid cell value in last FDTD iteration. To obtain the second part, we have to compute the discrete Laplacian:
\begin{align}
    \tilde{\nabla}^2p^n_{u,v,w}( \mathcal{F}_{u,v,w}) &= p^n_{u-1,v,w}( \mathcal{F}_{u,v,w}) +  p^n_{u+1,v,w}( \mathcal{F}_{u,v,w}) \nonumber\\
        &+ p^n_{u,v-1,w}( \mathcal{F}_{u,v,w}) +  p^n_{u,v+1,w}( \mathcal{F}_{u,v,w}) \nonumber\\
    &+ p^n_{u,v,w-1}( \mathcal{F}_{u,v,w}) +  p^n_{u,v,w+1}( \mathcal{F}_{u,v,w}) \nonumber\\
    &+ 6p^n_{u,v,w}( \mathcal{F}_{u,v,w}) \ .
\end{align}
The item \(6p^n_{u,v,w}( \mathcal{F}_{u,v,w}) \) can also be directly obtained from the grid cell value in last FDTD iteration. The other items have to be corrected from the grid cell value in last FDTD iteration. Here we take the computation process of \(p^n_{u-1,v,w}( \mathcal{F}_{u,v,w}) \) as an example. After the last iteration, we can obtain the far-field value of the neighbor cell \((u-1, v, w)\), that is, \(p^n_{u-1,v,w}( \mathcal{F}_{u-1,v,w})\). We can apply the correction from \(p^n_{u-1,v,w}( \mathcal{F}_{u-1,v,w})\) to \(p^n_{u-1,v,w}( \mathcal{F}_{u,v,w}) \)  as:
\begin{align}
    &p^n_{u-1,v,w}( \mathcal{F}_{u,v,w}) \nonumber \\
    &= p^n_{u-1,v,w}( \mathcal{F}_{u-1,v,w}) - p^n_{u-1,v,w}(\mathcal{F}_{u-1,v,w} \setminus \mathcal{F}_{u,v,w}) \nonumber \\
    & \quad + p^n_{u-1,v,w}(\mathcal{F}_{u,v,w} \setminus \mathcal{F}_{u-1,v,w} ) \ .
    \label{eq:laplacian_correction}
\end{align}
The illustration of the correction process is shown in \autoref{fig:fdtd_correction}. The number of boundary elements in set \(\mathcal{F}_{u-1,v,w} \setminus \mathcal{F}_{u,v,w} \) and \(\mathcal{F}_{u,v,w} \setminus \mathcal{F}_{u-1,v,w}  \) is usually small compared to the total boundary element number ($R_1 \ll N$) so that the computation of the correction items by definition in \autoref{eq:compute_U} is fast.

After the correction is applied to all items in the discrete Laplacian, the far-field value at cell \((u,v,w)\) is obtained by \autoref{eq:FDTD_far_update}.

The second-order ghost cell method~\cite{Wang2018} requires a time-consuming global equation solution due to the interdependence of ghost cell values. In contrast, the computation of all cells in our method is independent and can be implemented in parallel easily.

\subsection{Dirichlet Value Update}
\label{subsec:interpolate_far_field}
Dirichlet and Neumann values of all boundary elements are required to compute the items in the discrete Laplacian in FDTD, e.g. $p^n_{u-1,v,w}(\mathcal{F}_{u-1,v,w} \setminus \mathcal{F}_{u,v,w})$ in \autoref{eq:laplacian_correction}. The Neumann value is provided by the surface acceleration of the object, but the Dirichlet value is unknown. Solving the Dirichlet value from \autoref{eq:discretized_boundary_equation} is time-consuming due to the high computational complexity. In this paper, we use the far-field value in FDTD simulation to provide a fast approximation of the Dirichlet value.

In \autoref{eq:discretized_boundary_equation}, matrix \(\mathbf{D}_{0}\) represents the contribution of each boundary element to others in terms of Dirichlet value at the current time step. As the Dirichlet value of each boundary element at the current time step hasn't spread to affect other boundary elements, \(\mathbf{D}_{0}\) is nearly diagonal. Similarly, \(\mathbf{I}\) is also a diagonal matrix, making the left-hand matrix \(\left(\frac{1}{2} \mathbf{I} - \mathbf{D}_{0}\right)\) approximate to a diagonal matrix. Consequently, each row of \autoref{eq:discretized_boundary_equation} can be solved separately.

Here, let's consider the \(i\)-th row of \autoref{eq:discretized_boundary_equation} as an example. It provides an equation for solving the Dirichlet value of boundary element \(\Gamma_i\), which can be calculated as follows:
\begin{align}
    \Phi_{n}^i = \frac{1}{\alpha_i}\sum_{j=0}^{L} \left(\mathbf{D}_{j,i} \cdot \mathbf{\Phi}_{n-j} - \mathbf{V}_{j,i} \cdot  \mathbf{G}_{n-j} \right) \ ,
    \label{eq:tdbem_row_equation}
\end{align}
where \(\mathbf{D}_{j,i}\) and \(\mathbf{V}_{j,i}\) are vectors representing the \(i\)-th row of matrices \(\mathbf{D}_{j}\) and \(\mathbf{V}_{j}\), respectively. $\alpha_i$ denotes the \(i\)-th diagonal element in matrix \(\left(\frac{1}{2} \mathbf{I} - \mathbf{D}_{0}\right)\). It's worth noting that for convenience, we have included an additional term \(\mathbf{D}_{0,i} \cdot  \mathbf{\Phi}_{n}\); in this context, we enforce \( \mathbf{\Phi}_{n} = 0\) during the computation of right hand side of equation for correctness.

Vectors \(\mathbf{D}_{j,i}\)  and \(\mathbf{V}_{j,i}\) can be computed by Gaussian quadrature. Denote \(P\) Gaussian quadrature points \(x_1, ..., x_P\) in boundary element \(\Gamma_i\). Then \(\mathbf{D}_{j,i}\) can be computed as:
\begin{equation}
    \mathbf{D}_{j,i} = |\Gamma_i| \sum_{k=1}^P \omega_k \mathbf{d}_j(x_k) \ ,
\label{eq:D_gaussian_quad}
\end{equation}
where \(\omega_k\) is the quadrature weight for point \(x_k\) and $|\Gamma_i|$ is the area of boundary element $\Gamma_i$. \(\mathbf{V}_{j,i}\) can also be computed as:
\begin{equation}
    \mathbf{V}_{j,i} = |\Gamma_i| \sum_{k=1}^P \omega_k \mathbf{v}_j(x_k) \ .
\label{eq:V_gaussian_quad}
\end{equation}
By substituting \autoref{eq:D_gaussian_quad} and \autoref{eq:V_gaussian_quad} into \autoref{eq:tdbem_row_equation}, we can get: 
\begin{align}
    \Phi_{n}^i =  \frac{1}{\alpha_i}|\Gamma_i|  \sum_{k=1}^P  \omega_k \sum_{j=0}^{L} \left(\mathbf{d}_{j}(x_k) \cdot \mathbf{\Phi}_{n-j} - \mathbf{v}_{j}(x_k) \cdot  \mathbf{G}_{n-j} \right) \ ,
\end{align}
Then, by substituting \autoref{eq:discretized_potential_equation}, we can find that the Dirichlet value of the boundary elements is just the weighted sum of the pressure values of the quadrature points inside it:
\begin{equation}
    \Phi_{n}^i = \frac{1}{\alpha_i} |\Gamma_i|  \sum_{k=1}^P  \omega_k p^n_{x_k} \ .
\end{equation}

By denoting the set of boundary elements inside surrounding \(R_2 \times R_2 \times R_2\) grid cells as \(\mathbf{N}_i\) and the set of other boundary elements as \(\mathbf{F}_i\), the pressure values of the quadrature points can be decomposed into two parts:
\begin{align}
    p^n_{x_k} = p^n_{x_k}(\mathbf{N}_i) + p^n_{x_k}(\mathbf{F}_i)\ .
\end{align}
The near-field item \(p^n_{x_k}(\mathbf{N}_i)\) can be computed by definition. As the elements in \(\mathbf{N}_i\) are much fewer than the overall elements (\(R_2 \ll N\)), the computation is fast. The far-field item \(p^n_{x_k}(\mathbf{F}_i)\), which is smoother than the near-field item, can be approximated by interpolation of values of the surrounding grid cells around point \(x_k\). Specifically, if we denote the surrounding \(2^3\) grid cells as \((u_1, v_1, w_1)\) to \((u_8, v_8, w_8)\), the far-field item can be approximated as:

\begin{equation}
    p^n_{x_k}(\mathbf{F}_i) = \sum_{l=1}^8 \gamma_l p^n_{u_l, v_l, w_l} (\mathbf{F}_i) \ ,
\end{equation}
where \(\gamma_l\) is the trilinear interpolation weight. Finally, we only need to compute \(p^n_{u_l, v_l, w_l} (\mathbf{F}_i)\) based on the far-field value \(p^n_{u_l, v_l, w_l} (\mathcal{F}_{u_l, v_l, w_l})\) at cell \(u_l, v_l, w_l\) obtained from the FDTD iteration:
\begin{align}
    p^n_{u_l, v_l, w_l} (\mathbf{F}_i) &= p^n_{u_l, v_l, w_l} (\mathcal{F}_{u_l, v_l, w_l}) \nonumber \\
    &-p^n_{u_l, v_l, w_l} (\mathcal{F}_{u_l, v_l, w_l} \setminus \mathbf{F}_i) \nonumber \\
    &+p^n_{u_l, v_l, w_l} (\mathbf{F}_i \setminus \mathcal{F}_{u_l, v_l, w_l}) \ .
\end{align}
The correction process is similar to the process shown in \autoref{fig:fdtd_correction}.
As the number of boundary elements in sets $\mathcal{F}_{u_l, v_l, w_l} \setminus \mathbf{F}_i$ and $\mathbf{F}_i \setminus \mathcal{F}_{u_l, v_l, w_l}$ is usually small compared to the total boundary element number ($R_1, R_2 \ll N$), the computation is also fast.

\subsection{Temporal Dynamic Interface}
The ghost cell method can encounter the \textit{fresh-cell problem}~\cite{Mittal2005}, which results from the movement of the object interface and translation from ghost cell to fluid cell. The fresh-cell problem can result in audible pops in the generated audio if not handled properly~\cite{Wang2018}. Thus, the pressure of the fresh cell must be calculated by an expensive global solution~\cite{MITTAL2008} or estimated by local interpolation~\cite{Wang2018}. In contrast, our method seamlessly handles dynamic interfaces because all cells are air cells.

\subsection{Speedup Strategies}
In a scene, the number of mesh elements directly affects the computational speed of the entire algorithm. Specifically, within the overall process framework (as shown in Figure \autoref{fig:overview}), the time complexity of the steps "FDTD update for far-field" and "Interpolation of far-field" is proportional to the number of elements, while the time complexity of the step "Update near field from neighbor elements" is proportional to the square of the number of elements.

To enhance computational efficiency, we can reduce the number of boundary elements by remeshing with a slight compromise in accuracy. The Neumann data on the resampled mesh are interpolated from the initial mesh. It's worth noting that we define a boundary element in a cell when the center point of the boundary element lies within the grid cell. Therefore, we constrain the longest side of triangles in the mesh to be shorter than the grid cell size to prevent any boundary element from spanning more than two cells.

Convolution weights (e.g., $\mathbf{d}_j, \mathbf{v}_j$) are precomputed in our method, involving the scaled inverse discrete Fourier transform of size $N$. These weights remain constant in a static scene and are stored for efficient reuse. In cases of short local periods in dynamic scenes, where the object's speed is significantly slower than the speed of sound, convolution weights can be approximated as unchanging for efficiency.

\begin{figure}[h]
\centering
    \includegraphics[width=0.45\linewidth]{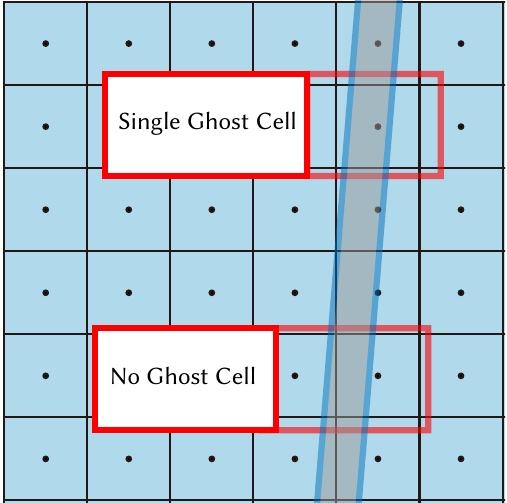}
    \caption{Illustration of ghost cell method for low grid resolution. For example, in the situation depicted in the figure, a single ghost cell cannot approximate the boundary condition with high accuracy. Moreover, the boundary condition information is entirely lost when no ghost cell can be located.}
\label{fig:noghostcell}
\end{figure}

\begin{figure*}[ht]
\centering
\includegraphics[width=1.0\linewidth]{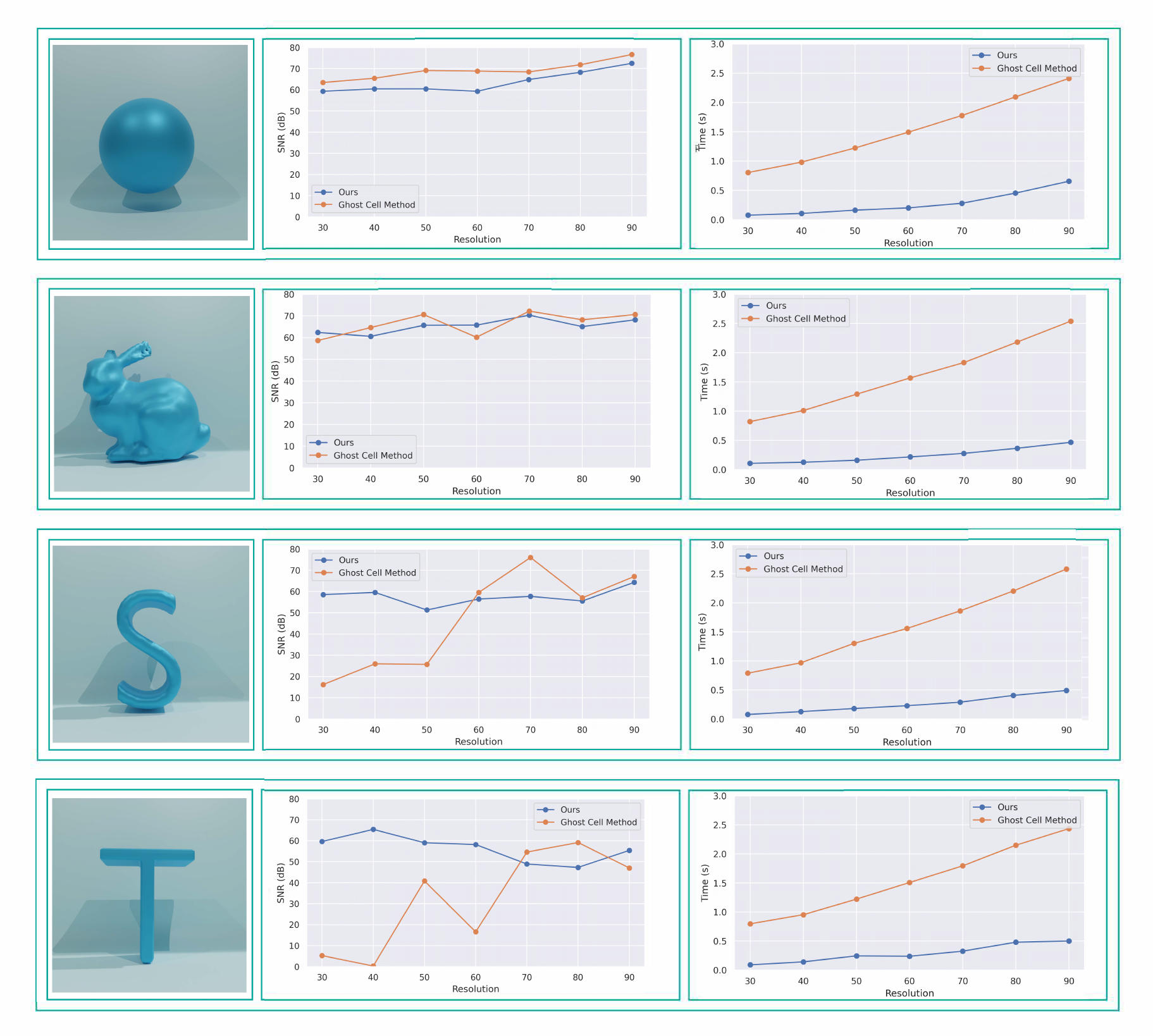}
\caption{
Our method vs. ghost cell-based method (second-order approximation)~\cite{Wang2018} in the monopole source test. Our method outperforms the ghost cell method across most of our tests. As we vary the resolution, our method consistently maintains high accuracy (measured by SNR) across different objects, while the ghost cell method experiences a drastic reduction in accuracy for low-resolution tests on the last two objects. Notably, our approach is always faster than the ghost cell method for all resolutions.
}
\label{fig:monopole}
\end{figure*}

\section{Experiments and Results}

\begin{table*}[ht]
\caption{Quantitative results for the monopole test. All values presented are averages across all testing scenarios. Our method outperforms the ghost cell-based baseline (\cite{Wang2018}) across all metrics when resolution is lower than 70 and achieves a remarkable speedup over the baseline at all resolutions. When the resolution exceeds 70, both methods deliver sufficiently accurate results, with the SNR fluctuating around 60.}
\label{table:monopole}
\vspace{0.1in}
\centering
\begin{tabular}{llccccccc}
\toprule
                            & Resolution & $30^3$    & $40^3$    & $50^3$    & $60^3$    & $70^3$    & $80^3$ & $90^3$\\ \hline
\multirow{2}{*}{Ghost Cell} & SNR        & 35.89 & 39.08 & 51.59 & 51.29 & \textbf{67.84}  & \textbf{64.05} & \textbf{65.34}\\
                            & Time (s)   & 0.80  & 0.96  & 1.26  & 1.52  & 1.81   & 2.16& 2.49\\ \hline
\multirow{2}{*}{Ours}       & SNR        & \textbf{59.98} & \textbf{61.49} & \textbf{59.11} & \textbf{59.93} & 60.45 & 59.03&65.11 \\
                            & Time (s)   & \textbf{0.09}  & \textbf{0.13}  & \textbf{0.19}  & \textbf{0.23}  & \textbf{0.29} &\textbf{0.42} & \textbf{0.53}\\ 
\bottomrule
\end{tabular}
\end{table*}

We conduct two quantitative experiments to showcase the power of our method and compare it with the previous baseline, the second-order ghost cell method~\cite{Wang2018} known for its high accuracy. First, we test our method and the baseline on a monopole scene with different meshes as boundaries (Sec. \ref{sec:monoploe_tes}). Next, we evaluate our method and the baseline in solving acoustic transfer (Sec. \ref{sec:acoustic_transfer_test}) of rigid body vibration. Finally, we utilize our method to synthesize sound for some complex sound radiation scenes, demonstrating its effectiveness in practical applications. Please refer to the supplementary video for detailed qualitative results presented both visually and acoustically.

In implementing both our method and the ghost cell method, Absorbing Boundary Conditions (ABC)~\cite{Wang2018} in FDTD boundary processing is incorporated to mitigate unnatural reflections at the boundary. The time step size is determined as large as possible under the Courant-Friedrichs-Lewy (CFL) condition. Both methods are implemented using CUDA for efficient computation. For a fair comparison, the most computationally intensive steps of the ghost cell method ~\cite{Wang2018}, including SVD and BiCGStab, are executed using cuSOLVER and cuSparse libraries for optimal performance. All tests are conducted on an RTX 3080Ti graphics card with an i7-8700 CPU. When the two boundary elements are not adjacent, the sample number for Gaussian quadrature points is set to 1. When the two boundary elements are adjacent, the sample number for Gaussian quadrature points is set to 3. The range of near-field for the grid cell is $R_1 = 3$ and the range of near-field for the boundary element is $R_2 = 4$. We use the history time step $L$ as 64.

\begin{figure*}[ht]
\centering
\subfigure[Spilling bowl]{
    \includegraphics[trim={5cm 0cm 5cm 0cm},clip,width=0.23\linewidth]{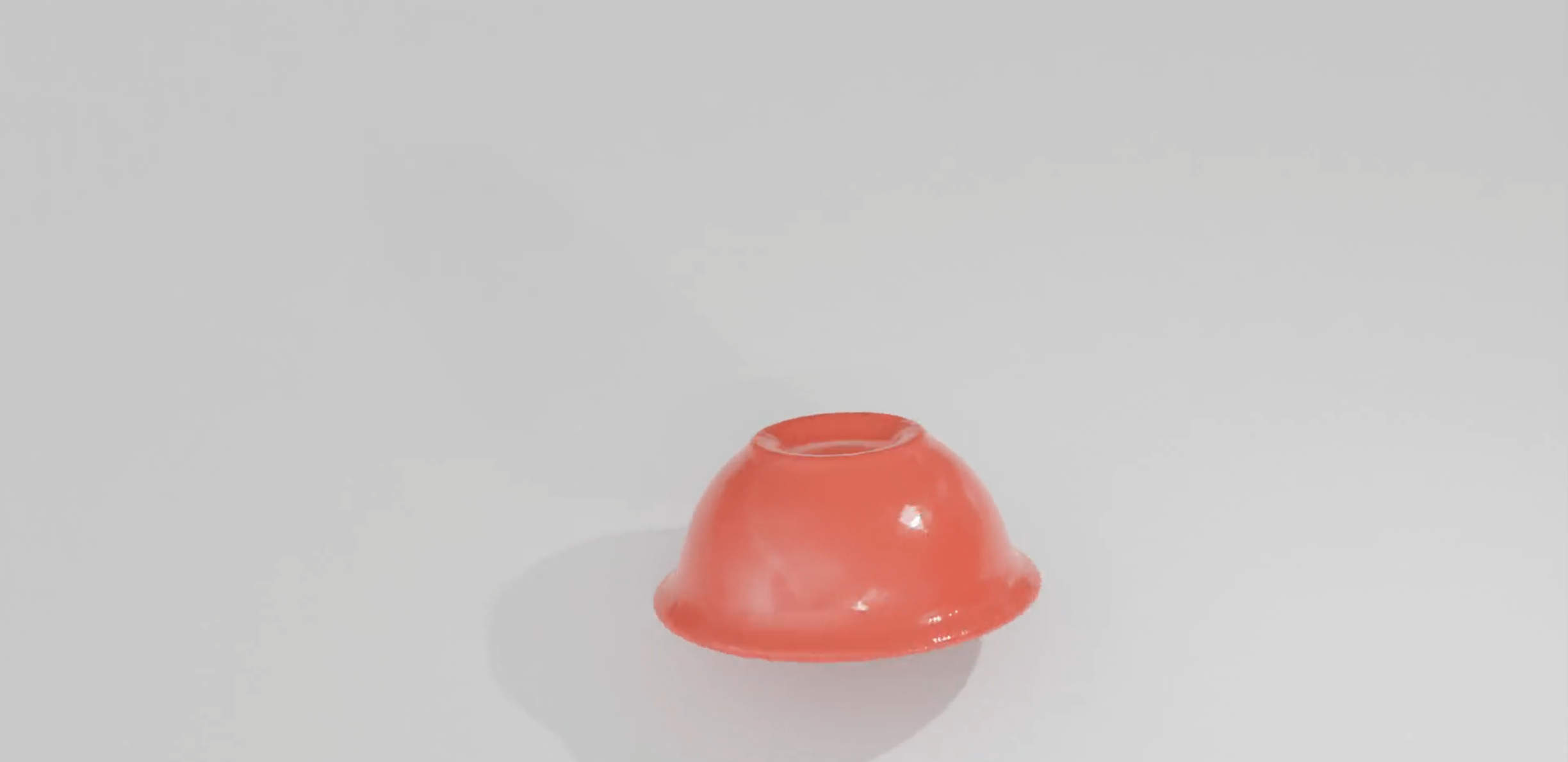}
    \label{fig:demo1}
    }
\subfigure[Pouring water]{
    \includegraphics[trim={5cm 0cm 5cm 0cm},clip,width=0.23\linewidth]{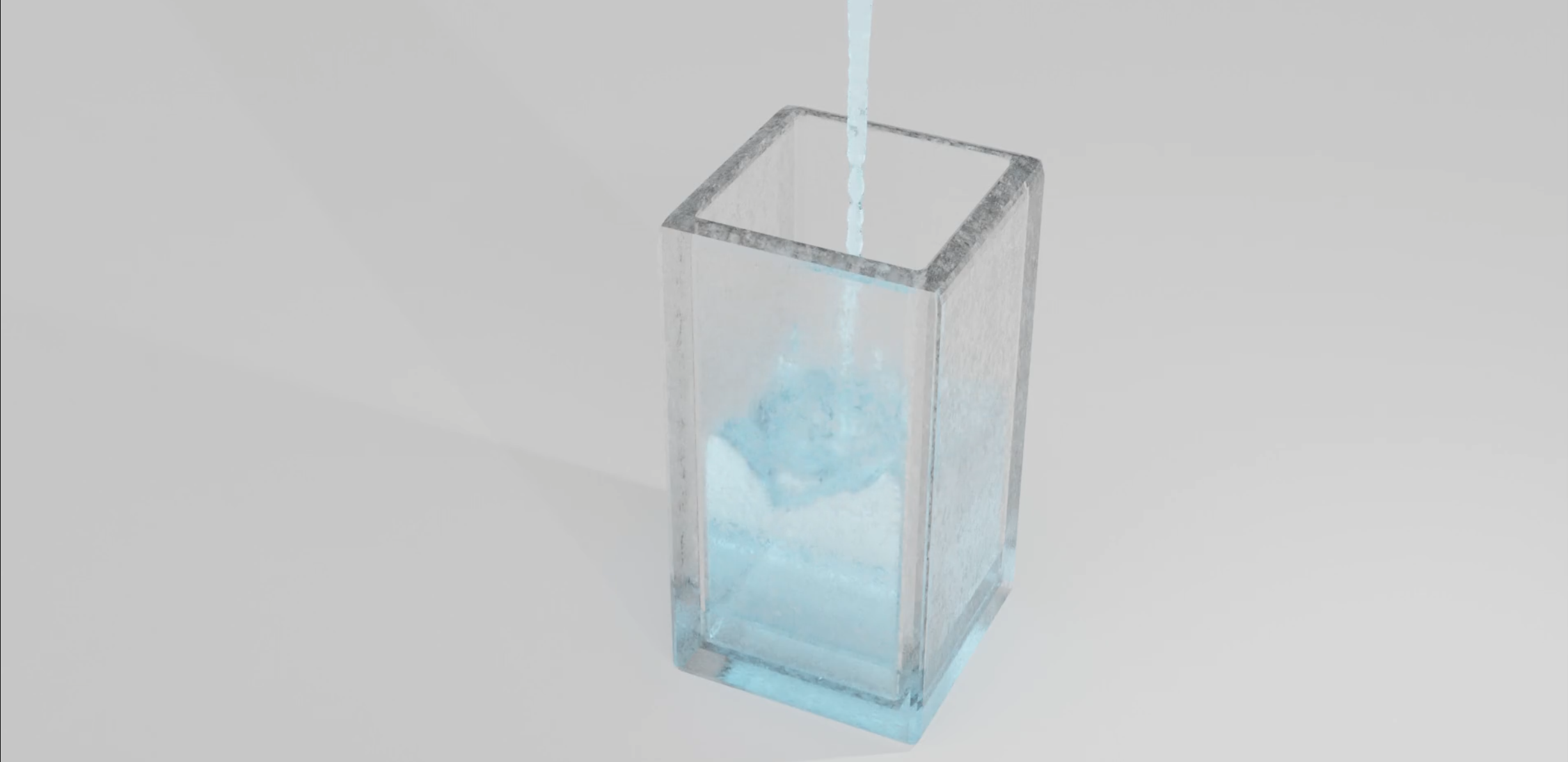}
    \label{fig:demo2}
    }
\subfigure[Ringing telephone]{
    \includegraphics[trim={5cm 0cm 5cm 0cm},clip,width=0.23\linewidth]{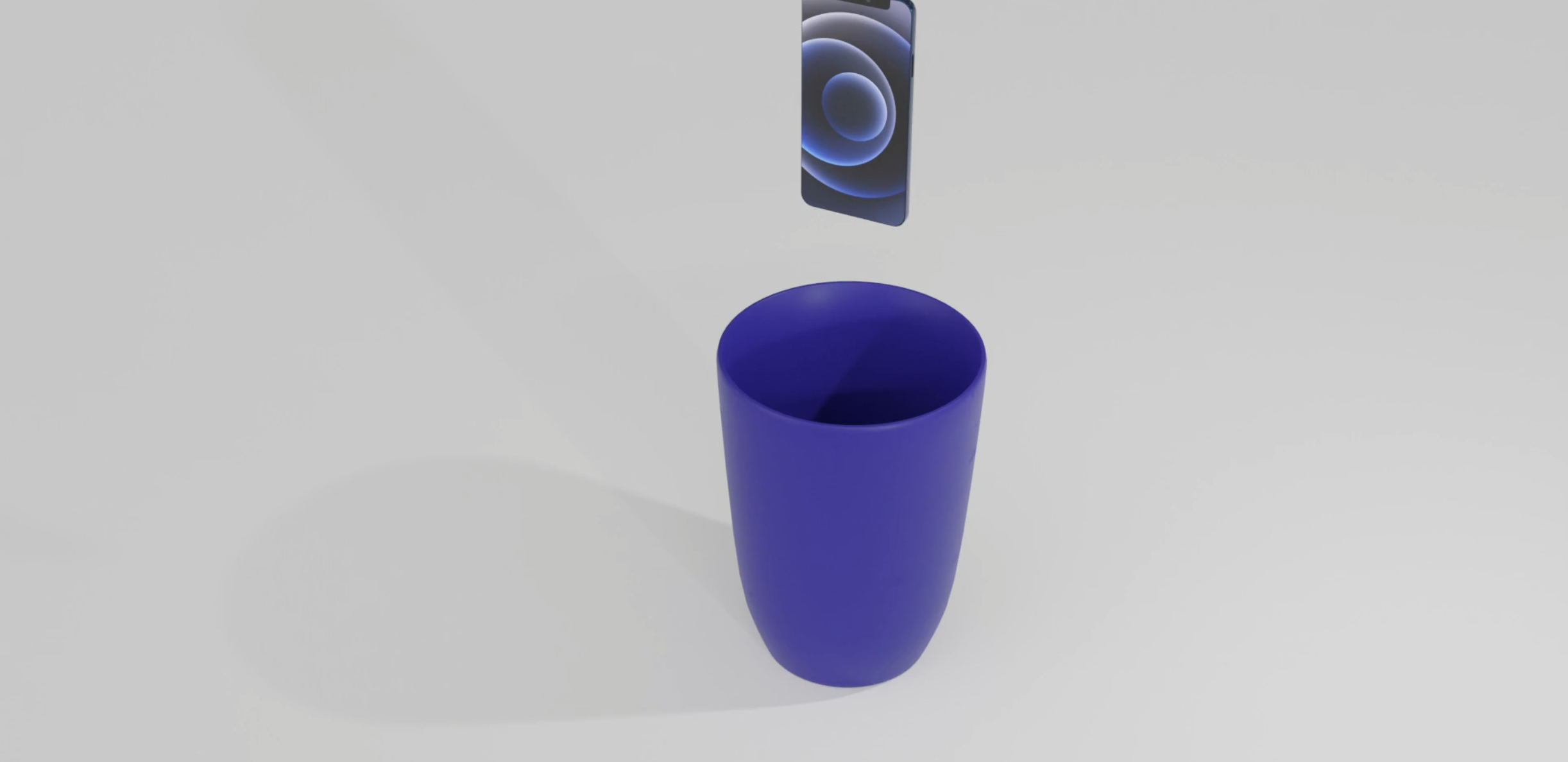}
    \label{fig:demo3}
    }
\subfigure[Speaker behind fan]{
    \includegraphics[trim={5cm 0cm 5cm 0cm},clip,width=0.23\linewidth]{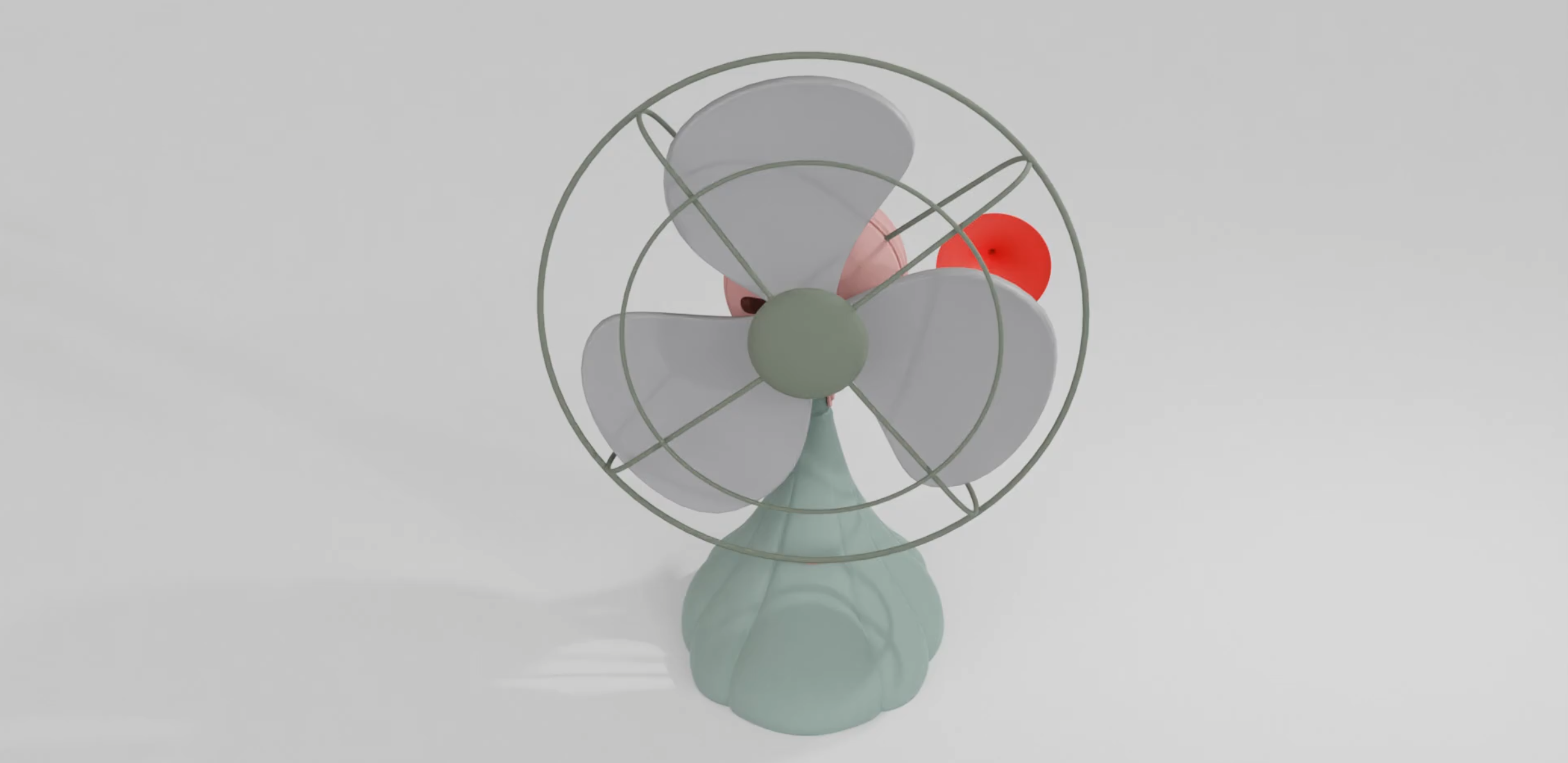}
    \label{fig:demo4}
    }
\caption{
Various natural scenarios of sound radiation by our approach, along with the sound demonstrations provided in the supplementary material.
}
\label{fig:demo}
\end{figure*}

\begin{figure*}[ht]
\centering
\includegraphics[width=0.98\linewidth]{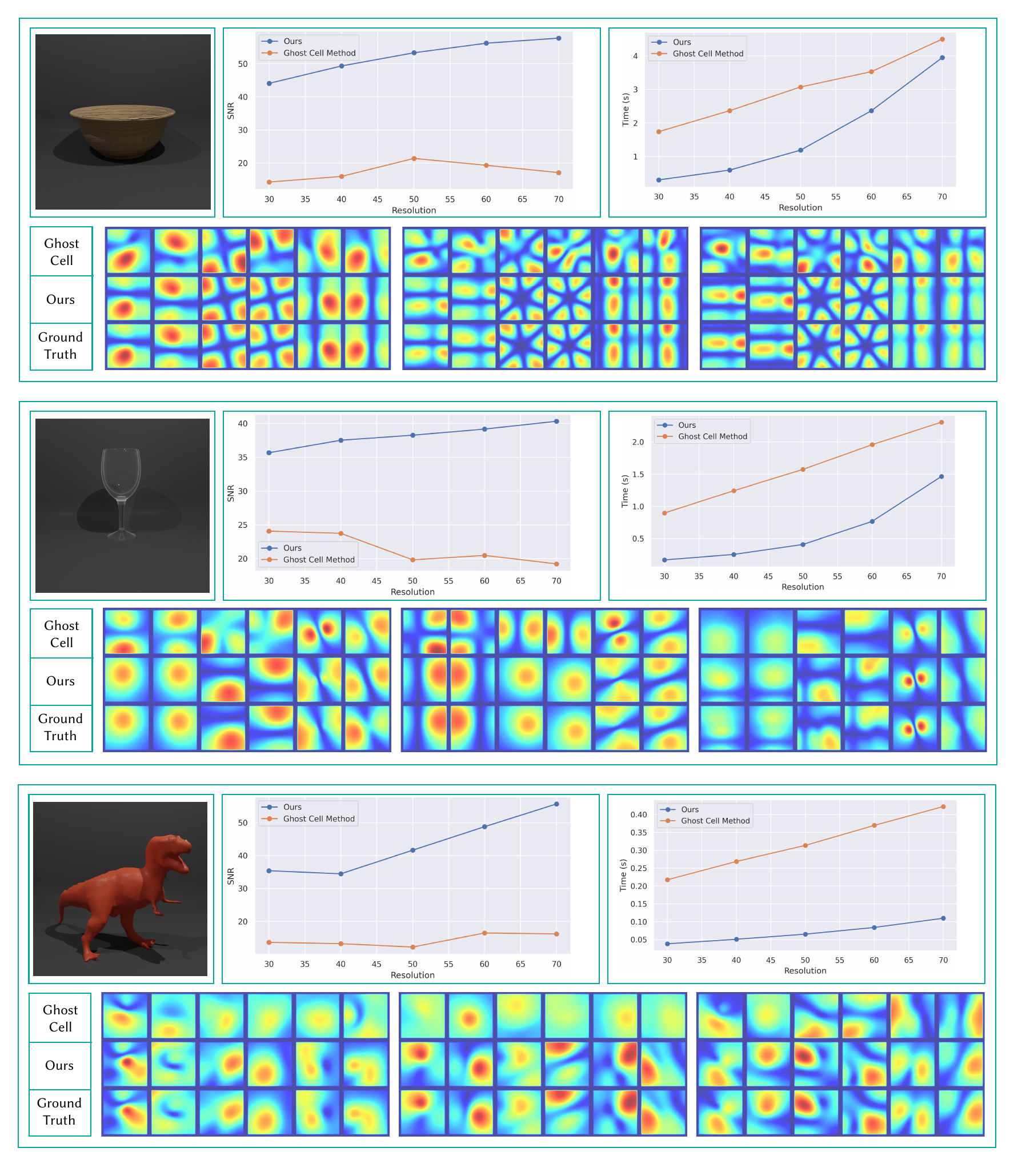}
\caption{
Comparison of the ghost cell-based method (second-order approximation)~\cite{Wang2018} and our approach in the acoustic transfer test. These tests were conducted with complex meshes and intricate surface Neumann data. The bottom images visualize the FFAT maps for modes 1, 2, and 3, displaying the six unfolded faces of the bounding box for each mode, with a $70\times 70\times 70$ grid resolution. Our method outperforms the ghost cell method in terms of speed and accuracy for all tested objects. When varying the resolution, our method consistently maintains high accuracy, as measured by SNR, across various grid sizes and different objects. The visualized FFAT map also illustrates that our method produces results much closer to the ground truth than the ghost cell method at resolution 70.
}
\label{fig:acoustic}
\end{figure*}

\subsection{Monopole Test}
\label{sec:monoploe_tes}
The monopole sound source serves as the unit test of the multipole field used for testing the sound radiation solver~\cite{sigsoundcourse}. We place a unit multipole source at the center of the bounding box of a closed mesh. The monopole source generates a pressure field at position $\mathbf{x}$ defined as:
\begin{equation}
    p(\mathbf{x}) = \frac{e^{-ik|\mathbf{x} - \mathbf{c}|}}{4 \pi |\mathbf{x} - \mathbf{c}|}\ ,
    \label{eq:monopole_test}
\end{equation}
where $\mathbf{c}$ is the source location and $k$ is the wave number. During simulation, the time-domain term \(e^{i\omega t}\) is included in the time-domain signal. We obtain the Neumann boundary condition data by numerically evaluating $\partial_n p(\mathbf{x})$ for each boundary element, and then apply the Neumann data to the solver. Subsequently, we sample the pressure values on the surface of a bounding box and compare these values with the ground truth computed by \autoref{eq:monopole_test}.

A sphere, a bunny, and two letter-shaped meshes are used as the typical test objects. The simulation domain spans 0.7 meters, and the mesh size is scaled to 1/6 of this domain to minimize boundary effects. A monopole source at 1000 Hz is employed. The bounding boxes for sampling and evaluation are generated by extending the object's bounding box by factors of 2.4, 3.0, 3.6, and 4.2 for near-field checks. The sample resolution on the surface of the bounding box is the same as the grid resolution. The signal-to-noise ratio (SNR) in decibels (dB) is used as a metric for comparison against the ground truth.

Specifically, if we totally sample $n$ points on the bounding boxes,  denote the solved pressure values of these points as $\tilde{\mathbf{p}} = (\tilde{p_1}, \tilde{p_2}, ...., \tilde{p_n})$ and the corresponding ground truth as $\mathbf{p} = (p_1, p_2, ..., p_n)$. The SNR is calculated using the following equation:
\begin{equation}
    \text{SNR} = 10 \log \left( \frac{\sum_{i=1}^n p_i^2}{\sum_{i = 1}^n (\tilde{p_i} - p_i)^2} \right)\ .
\end{equation}

Different grid resolutions are tested for comparison between our method and the ghost cell-based method. We record the SNR of pressure values and the time cost for each resolution. The corresponding line plots for each object are shown in \autoref{fig:monopole}. Our method consistently required less time cost than ghost cell-based method for each grid resolution. This is because the ghost cell-based method necessitates a costly global solution of linear equations for each step to ensure second-order boundary approximation accuracy~\cite{Wang2018}, whereas our method relies only on quick local operations for boundary processing. Our method maintained a high SNR even when a low-resolution grid was used for all objects, while the ghost cell-based method faced drastic drops in accuracy with low-resolution grids. We also presented the mean SNR and time over these objects in~\autoref{table:monopole}. This table illustrates that our method produced more accurate results than the ghost cell-based method for resolutions ranging from 30 to 60. Our method can consistently achieve high accuracy ($\sim$60 SNR) even at low resolutions in 0.09s, while the ghost cell-based method requires more than 1.52s to achieve the same level of accuracy, achieving a $\sim 17 \times$ speedup.

To explain why the ghost cell-based method faces drastic accuracy reduction for low grid resolution, we present an example in \autoref{fig:noghostcell}. The ghost cell must be a solid cell with at least one neighboring air cell. When the grid is large compared to the local geometry shape, the ghost cell can be hard to locate or may not be located at all. The absence of a ghost cell leads to the loss of boundary conditions around it, resulting in a reduction in accuracy. Our method provides stable and accurate solutions even when using larger grid cell sizes (as seen in the results for letter-shaped meshes in \autoref{fig:monopole}), as it does not rely on ghost cells.

\subsection{Acoustic Transfer Test}
\label{sec:acoustic_transfer_test}
In the monopole test, the Neumann condition is smooth around the boundaries. To assess the performance of both our method and the ghost cell-based method on more complex Neumann conditions, we employ them to solve the acoustic transfer function of a vibrating rigid body. The surface Neumann data is obtained from modal analysis on the rigid body. We utilize the FFAT map~\cite{shell, KleinPAT} to represent the acoustic transfer function. The ground truth of the FFAT map is calculated using the frequency-domain boundary element method (BEM) with the \textit{bempp} library~\cite{bempp}.

For these experiments, we select three complex objects (a wooden bowl, glass, and plastic dinosaur) and solve the FFAT map for their first five vibration modes. The simulation domain size is set to 2.5 times the bounding box to encompass more cells around the objects. We conduct the experiments using various grid resolutions.
 
As shown in \autoref{fig:acoustic}, our method outperforms the ghost cell-based method in terms of both SNR and computational efficiency in all tests. When varying the resolution, our method consistently maintains high accuracy (measured by SNR) across various grid sizes and different objects. It's worth noting that for larger objects, like the dinosaur, the grid cell size is larger. This leads to a larger time step size under the CFL condition and fewer steps needed within the same simulation time.

The figure also includes visualizations of the FFAT maps for modes 1, 2, and 3 from experiments using a $70\times 70\times 70$ grid resolution. These visualizations display the six unfolded faces of the bounding box for each mode, demonstrating that the results of our method are closer to the ground truth.

\subsection{Natural Scenarios}
We showcase dynamic sound results generated by our method, demonstrating its ability to produce suitable sound effects acrossCASAXR diverse natural scenarios, as shown in \autoref{fig:demo}. Specifically, we use Blender to create and render the scenes, employing our method to solve the sound radiation. We then select a listener position for audio synthesis. The following scenarios are included:

\begin{itemize}
    \item Spilling bowl: Our approach synthesizes the resonance produced by a bowl spilling on the floor. The surface Neumann data is obtained through rigid body modal analysis. The dynamic changes in the angle between the bowl and the ground result in corresponding variations in the emitted sound.
    \item Pouring water: Our approach synthesizes the time-varying resonance caused by water rising in a container. We assume the fluid surface to be a plane, with the Neumann data obtained from a water sound audio source~\cite{Langlois2016}. Both the water surface and the Neumann data exhibit temporal variations.
    \item Ringing telephone: The resonance of a ringing telephone undergoes a change as it shifts from an outdoor environment to being immersed in a cup. Our approach accurately synthesizes this variation in resonance. Changes in the geometric relationship between the mobile phone and the cup will cause corresponding alterations in the sound.
    \item Loudspeaker behind rotating fan: A loudspeaker positioned behind a rotating fan generates a distinctive `robotic' voice. The pitch and character of this voice vary in response to changes in the fan's rotational speed, with different angular velocities producing different vocal effects.
\end{itemize}

For additional information, we recommend watching the accompanying video. Our solver takes 3 to 20 minutes to simulate these scenarios on a consumer graphics card, while the ghost cell-based method takes hours to days for similar scenarios on a Google Cloud Compute platform with hundreds of cores~\cite{Wang2018}.

\section{Conclusion, Limitations, and Future Work}

We have achieved a significant enhancement in radiation simulation, particularly in handling complex boundary conditions by getting rid of the ghost cells. Our hybrid approach effectively marries the time-domain boundary element method for near-field sound radiation calculations with the finite-difference time-domain method for far-field radiation. This results in a highly efficient and accurate computational solution, regardless of the resolution of the grid. However, there is still room for improvement. Future research could include a more thorough theoretical analysis of the errors introduced by our method and the implementation of methods to correct them with increased precision. Our remeshing strategy is rather straightforward, and the use of a frequency-adaptive remeshing technique~\cite{Li2016} would allow for higher precision in the reproduction of high-frequency sounds.

Our approach has the potential to drive advancements in grid-based physical simulation. For instance, our approach may help improve PPPM~\cite{ANDERSON1986,Zhang2014} in fluid dynamics for higher accuracy. Given its ability to tolerate poor meshes and large grid cell sizes, our approach holds promise for achieving near real-time sound radiation simulation. Additionally, our approach can be combined with the KleinPAT~\cite{KleinPAT} method to achieve faster and more accurate acoustic transfer map calculations. However, these areas still require further investigation.

\bibliographystyle{ACM-Reference-Format}
\bibliography{SoundRadiation}

@article{smith1992,
 ISSN = {01489267, 15315169},
 URL = {http://www.jstor.org/stable/3680470},
 author = {Julius O. Smith},
 journal = {Computer Music Journal},
 number = {4},
 pages = {74--91},
 publisher = {The MIT Press},
 title = {Physical Modeling Using Digital Waveguides},
 urldate = {2023-01-13},
 volume = {16},
 year = {1992}
}

@inproceedings{Gaver1993,
author = {Gaver, William W.},
title = {Synthesizing Auditory Icons},
year = {1993},
isbn = {0897915755},
publisher = {Association for Computing Machinery},
address = {New York, NY, USA},
url = {https://doi.org/10.1145/169059.169184},
doi = {10.1145/169059.169184},
booktitle = {Proceedings of the INTERACT '93 and CHI '93 Conference on Human Factors in Computing Systems},
pages = {228–235},
numpages = {8},
location = {Amsterdam, The Netherlands},
series = {CHI '93}
}

@ARTICLE{cook2002,
  author={Cook, P.R.},
  journal={IEEE Computer Graphics and Applications}, 
  title={Sound production and modeling}, 
  year={2002},
  volume={22},
  number={4},
  pages={23-27},
  doi={10.1109/MCG.2002.1016695}
}

@misc{DineshSurvey,
  doi = {10.48550/ARXIV.2011.05538},
  
  url = {https://arxiv.org/abs/2011.05538},
  
  author = {Liu, Shiguang and Manocha, Dinesh},
  
  title = {Sound Synthesis, Propagation, and Rendering: A Survey},
  
  publisher = {arXiv},
  year = {2020},
  copyright = {Creative Commons Attribution 4.0 International}
}

@inproceedings{OBrien2002,
author = {O'Brien, James F. and Shen, Chen and Gatchalian, Christine M.},
title = {Synthesizing Sounds from Rigid-Body Simulations},
year = {2002},
isbn = {1581135734},
publisher = {Association for Computing Machinery},
address = {New York, NY, USA},
booktitle = {Proceedings of the 2002 ACM SIGGRAPH/Eurographics Symposium on Computer Animation},
pages = {175–181},
numpages = {7},
location = {San Antonio, Texas},
series = {SCA '02}
}

@inproceedings{Raghuvanshi2006,
author = {Raghuvanshi, Nikunj and Lin, Ming C.},
title = {Interactive Sound Synthesis for Large Scale Environments},
year = {2006},
isbn = {159593295X},
publisher = {Association for Computing Machinery},
address = {New York, NY, USA},
url = {https://doi.org/10.1145/1111411.1111429},
doi = {10.1145/1111411.1111429},
booktitle = {Proceedings of the 2006 Symposium on Interactive 3D Graphics and Games},
pages = {101–108},
numpages = {8},
location = {Redwood City, California},
series = {I3D '06}
}

@inproceedings{van2001,
author = {van den Doel, Kees and Kry, Paul G. and Pai, Dinesh K.},
title = {FoleyAutomatic: Physically-Based Sound Effects for Interactive Simulation and Animation},
year = {2001},
isbn = {158113374X},
publisher = {Association for Computing Machinery},
address = {New York, NY, USA},
url = {https://doi.org/10.1145/383259.383322},
doi = {10.1145/383259.383322},
booktitle = {Proceedings of the 28th Annual Conference on Computer Graphics and Interactive Techniques},
pages = {537–544},
numpages = {8},
series = {SIGGRAPH '01}
}

@article{Chadwick2012,
author = {Chadwick, Jeffrey N. and Zheng, Changxi and James, Doug L.},
title = {Precomputed Acceleration Noise for Improved Rigid-Body Sound},
year = {2012},
issue_date = {July 2012},
publisher = {Association for Computing Machinery},
address = {New York, NY, USA},
volume = {31},
number = {4},
issn = {0730-0301},
url = {https://doi.org/10.1145/2185520.2185599},
doi = {10.1145/2185520.2185599},
journal = {ACM Trans. Graph.},
month = {jul},
articleno = {103},
numpages = {9}
}

@article{contact,
author = {Zheng, Changxi and James, Doug L.},
title = {Toward High-Quality Modal Contact Sound},
year = {2011},
issue_date = {July 2011},
publisher = {Association for Computing Machinery},
address = {New York, NY, USA},
volume = {30},
number = {4},
issn = {0730-0301},
url = {https://doi.org/10.1145/2010324.1964933},
doi = {10.1145/2010324.1964933},
journal = {ACM Trans. Graph.},
month = {jul},
articleno = {38},
numpages = {12}
}

@article{Ren2013,
author = {Ren, Zhimin and Yeh, Hengchin and Lin, Ming C.},
title = {Example-Guided Physically Based Modal Sound Synthesis},
year = {2013},
issue_date = {January 2013},
publisher = {Association for Computing Machinery},
address = {New York, NY, USA},
volume = {32},
number = {1},
issn = {0730-0301},
url = {https://doi.org/10.1145/2421636.2421637},
doi = {10.1145/2421636.2421637},
journal = {ACM Trans. Graph.},
month = {feb},
articleno = {1},
numpages = {16}
}

@article{NeuralSound,
author = {Jin, Xutong and Li, Sheng and Wang, Guoping and Manocha, Dinesh},
title = {NeuralSound: Learning-Based Modal Sound Synthesis with Acoustic Transfer},
year = {2022},
issue_date = {July 2022},
publisher = {Association for Computing Machinery},
address = {New York, NY, USA},
volume = {41},
number = {4},
issn = {0730-0301},
url = {https://doi.org/10.1145/3528223.3530184},
doi = {10.1145/3528223.3530184},
journal = {ACM Trans. Graph.},
month = {jul},
articleno = {121},
numpages = {15}
}

@inproceedings{DeepModal,
author = {Jin, Xutong and Li, Sheng and Qu, Tianshu and Manocha, Dinesh and Wang, Guoping},
title = {Deep-Modal: Real-Time Impact Sound Synthesis for Arbitrary Shapes},
year = {2020},
isbn = {9781450379885},
publisher = {Association for Computing Machinery},
address = {New York, NY, USA},
url = {https://doi.org/10.1145/3394171.3413572},
doi = {10.1145/3394171.3413572},
booktitle = {Proceedings of the 28th ACM International Conference on Multimedia},
pages = {1171–1179},
numpages = {9},
location = {Seattle, WA, USA},
series = {MM '20}
}

@article{HarmonicShell,
author = {Chadwick, Jeffrey N. and An, Steven S. and James, Doug L.},
title = {Harmonic Shells: A Practical Nonlinear Sound Model for near-Rigid Thin Shells},
year = {2009},
issue_date = {December 2009},
publisher = {Association for Computing Machinery},
address = {New York, NY, USA},
volume = {28},
number = {5},
issn = {0730-0301},
url = {https://doi.org/10.1145/1618452.1618465},
doi = {10.1145/1618452.1618465},
journal = {ACM Trans. Graph.},
month = {dec},
pages = {1–10},
numpages = {10}
}

@article{Minnaert1933,
author = { M.   Minnaert   Sc.D. },
title = {XVI. On musical air-bubbles and the sounds of running water},
journal = {The London, Edinburgh, and Dublin Philosophical Magazine and Journal of Science},
volume = {16},
number = {104},
pages = {235-248},
year  = {1933},
publisher = {Taylor & Francis},
doi = {10.1080/14786443309462277},
URL = { https://doi.org/10.1080/14786443309462277},
eprint = { https://doi.org/10.1080/14786443309462277}
}

@article{Moss2010,
author = {Moss, William and Yeh, Hengchin and Hong, Jeong-Mo and Lin, Ming C. and Manocha, Dinesh},
title = {Sounding Liquids: Automatic Sound Synthesis from Fluid Simulation},
year = {2010},
issue_date = {June 2010},
publisher = {Association for Computing Machinery},
address = {New York, NY, USA},
volume = {29},
number = {3},
issn = {0730-0301},
url = {https://doi.org/10.1145/1805964.1805965},
doi = {10.1145/1805964.1805965},
journal = {ACM Trans. Graph.},
month = {jul},
articleno = {21},
numpages = {13}
}

@article{HarmonicFluids,
author = {Zheng, Changxi and James, Doug L.},
title = {Harmonic Fluids},
year = {2009},
issue_date = {August 2009},
publisher = {Association for Computing Machinery},
address = {New York, NY, USA},
volume = {28},
number = {3},
issn = {0730-0301},
url = {https://doi.org/10.1145/1531326.1531343},
doi = {10.1145/1531326.1531343},
journal = {ACM Trans. Graph.},
month = {jul},
articleno = {37},
numpages = {12}
}

@Article{Kirkup2019,
AUTHOR = {Kirkup, Stephen},
TITLE = {The Boundary Element Method in Acoustics: A Survey},
JOURNAL = {Applied Sciences},
VOLUME = {9},
YEAR = {2019},
NUMBER = {8},
ARTICLE-NUMBER = {1642},
URL = {https://www.mdpi.com/2076-3417/9/8/1642},
ISSN = {2076-3417},
DOI = {10.3390/app9081642}
}

@article{PAT,
author = {James, Doug L. and Barbi\v{c}, Jernej and Pai, Dinesh K.},
title = {Precomputed Acoustic Transfer: Output-Sensitive, Accurate Sound Generation for Geometrically Complex Vibration Sources},
year = {2006},
issue_date = {July 2006},
publisher = {Association for Computing Machinery},
address = {New York, NY, USA},
volume = {25},
number = {3},
issn = {0730-0301},
url = {https://doi.org/10.1145/1141911.1141983},
doi = {10.1145/1141911.1141983},
journal = {ACM Trans. Graph.},
month = {jul},
pages = {987–995},
numpages = {9}
}

@article{Langlois2014,
author = {Langlois, Timothy R. and An, Steven S. and Jin, Kelvin K. and James, Doug L.},
title = {Eigenmode Compression for Modal Sound Models},
year = {2014},
issue_date = {July 2014},
publisher = {Association for Computing Machinery},
address = {New York, NY, USA},
volume = {33},
number = {4},
issn = {0730-0301},
url = {https://doi.org/10.1145/2601097.2601177},
doi = {10.1145/2601097.2601177},
journal = {ACM Trans. Graph.},
month = {jul},
articleno = {40},
numpages = {9}
}

@ARTICLE{Rungta2016,
  author={Rungta, Atul and Schissler, Carl and Mehra, Ravish and Malloy, Chris and Lin, Ming and Manocha, Dinesh},
  journal={IEEE Transactions on Visualization and Computer Graphics}, 
  title={SynCoPation: Interactive Synthesis-Coupled Sound Propagation}, 
  year={2016},
  volume={22},
  number={4},
  pages={1346-1355},
  doi={10.1109/TVCG.2016.2518421}}

@article{Zheng2010,
author = {Zheng, Changxi and James, Doug L.},
title = {Rigid-Body Fracture Sound with Precomputed Soundbanks},
year = {2010},
issue_date = {July 2010},
publisher = {Association for Computing Machinery},
address = {New York, NY, USA},
volume = {29},
number = {4},
issn = {0730-0301},
url = {https://doi.org/10.1145/1778765.1778806},
doi = {10.1145/1778765.1778806},
journal = {ACM Trans. Graph.},
month = {jul},
articleno = {69},
numpages = {13}
}

@article{KleinPAT,
author = {Wang, Jui-Hsien and James, Doug L.},
title = {KleinPAT: Optimal Mode Conflation for Time-Domain Precomputation of Acoustic Transfer},
year = {2019},
issue_date = {August 2019},
publisher = {Association for Computing Machinery},
address = {New York, NY, USA},
volume = {38},
number = {4},
issn = {0730-0301},
url = {https://doi.org/10.1145/3306346.3322976},
doi = {10.1145/3306346.3322976},
journal = {ACM Trans. Graph.},
month = {jul},
articleno = {122},
numpages = {12}
}

@book{larsson2003,
  title={Partial differential equations with numerical methods},
  author={Larsson, Stig and Thom{\'e}e, Vidar},
  volume={45},
  year={2003},
  publisher={Springer}
}

@article{botteldooren1995,
  title={Finite-difference time-domain simulation of low-frequency room acoustic problems},
  author={Botteldooren, Dick},
  journal={The Journal of the Acoustical Society of America},
  volume={98},
  number={6},
  pages={3302--3308},
  year={1995},
  publisher={Acoustical Society of America}
}

@ARTICLE{Bilbao2013,
  author={Bilbao, Stefan},
  journal={IEEE Transactions on Audio, Speech, and Language Processing}, 
  title={Modeling of Complex Geometries and Boundary Conditions in Finite Difference/Finite Volume Time Domain Room Acoustics Simulation}, 
  year={2013},
  volume={21},
  number={7},
  pages={1524-1533},
  doi={10.1109/TASL.2013.2256897}}

@book{bilbao2009,
  title={Numerical sound synthesis: finite difference schemes and simulation in musical acoustics},
  author={Bilbao, Stefan},
  year={2009},
  publisher={John Wiley \& Sons}
}

@inproceedings{Micikevicius2009,
author = {Micikevicius, Paulius},
title = {3D Finite Difference Computation on GPUs Using CUDA},
year = {2009},
isbn = {9781605585178},
publisher = {Association for Computing Machinery},
address = {New York, NY, USA},
url = {https://doi.org/10.1145/1513895.1513905},
doi = {10.1145/1513895.1513905},
booktitle = {Proceedings of 2nd Workshop on General Purpose Processing on Graphics Processing Units},
pages = {79–84},
numpages = {6},
location = {Washington, D.C., USA},
series = {GPGPU-2}
}

@article{MEHRA201283,
title = {An efficient GPU-based time domain solver for the acoustic wave equation},
journal = {Applied Acoustics},
volume = {73},
number = {2},
pages = {83-94},
year = {2012},
issn = {0003-682X},
doi = {https://doi.org/10.1016/j.apacoust.2011.05.012},
url = {https://www.sciencedirect.com/science/article/pii/S0003682X11001605},
author = {Ravish Mehra and Nikunj Raghuvanshi and Lauri Savioja and Ming C. Lin and Dinesh Manocha},
}

@article{Allen2015,
author = {Allen, Andrew and Raghuvanshi, Nikunj},
title = {Aerophones in Flatland: Interactive Wave Simulation of Wind Instruments},
year = {2015},
issue_date = {August 2015},
publisher = {Association for Computing Machinery},
address = {New York, NY, USA},
volume = {34},
number = {4},
issn = {0730-0301},
url = {https://doi.org/10.1145/2767001},
doi = {10.1145/2767001},
journal = {ACM Trans. Graph.},
month = {jul},
articleno = {134},
numpages = {11}
}

@article{Wang2018,
author = {Wang, Jui-Hsien and Qu, Ante and Langlois, Timothy R. and James, Doug L.},
title = {Toward Wave-Based Sound Synthesis for Computer Animation},
year = {2018},
issue_date = {August 2018},
publisher = {Association for Computing Machinery},
address = {New York, NY, USA},
volume = {37},
number = {4},
issn = {0730-0301},
url = {https://doi.org/10.1145/3197517.3201318},
doi = {10.1145/3197517.3201318},
journal = {ACM Trans. Graph.},
month = {jul},
articleno = {109},
numpages = {16}
}

@incollection{langer2008,
  title={Time domain boundary element method},
  author={Langer, Sabine and Schanz, Martin},
  booktitle={Computational Acoustics of Noise Propagation in Fluids-Finite and Boundary Element Methods},
  pages={495--516},
  year={2008},
  publisher={Springer}
}

@article{Lubich1994,
author = {Lubich, Ch.},
title = {On the Multistep Time Discretization of Linear Initial-Boundary Value Problems and Their Boundary Integral Equations},
year = {1994},
issue_date = {April 1994},
publisher = {Springer-Verlag},
address = {Berlin, Heidelberg},
volume = {67},
number = {3},
issn = {0029-599X},
url = {https://doi.org/10.1007/s002110050033},
doi = {10.1007/s002110050033},
journal = {Numer. Math.},
month = {apr},
pages = {365–389},
numpages = {25}
}

@article{Banjai2010,
author = {Banjai, Lehel},
title = {Multistep and Multistage Convolution Quadrature for the Wave Equation: Algorithms and Experiments},
journal = {SIAM Journal on Scientific Computing},
volume = {32},
number = {5},
pages = {2964-2994},
year = {2010},
doi = {10.1137/090775981},
URL = { https://doi.org/10.1137/090775981},
eprint = { https://doi.org/10.1137/090775981}
}

@article{Banjai2008,
author = {Banjai, L. and Sauter, S.},
title = {Rapid Solution of the Wave Equation in Unbounded Domains},
year = {2008},
issue_date = {October 2008},
publisher = {Society for Industrial and Applied Mathematics},
address = {USA},
volume = {47},
number = {1},
issn = {0036-1429},
url = {https://doi.org/10.1137/070690754},
doi = {10.1137/070690754},
journal = {SIAM J. Numer. Anal.},
month = {oct},
pages = {227–249},
numpages = {23}
}

@book{liu2009,
  title={Fast multipole boundary element method: theory and applications in engineering},
  author={Liu, Yijun},
  year={2009},
  publisher={Cambridge university press}
}

@article{COUET1981,
title = {Simulation of three-dimensional incompressible flows with a vortex-in-cell method},
journal = {Journal of Computational Physics},
volume = {39},
number = {2},
pages = {305-328},
year = {1981},
issn = {0021-9991},
doi = {https://doi.org/10.1016/0021-9991(81)90154-6},
url = {https://www.sciencedirect.com/science/article/pii/0021999181901546},
author = {Benoit Couët and Oscar Buneman and Anthony Leonard},
}

@article{ANDERSON1986,
title = {A method of local corrections for computing the velocity field due to a distribution of vortex blobs},
journal = {Journal of Computational Physics},
volume = {62},
number = {1},
pages = {111-123},
year = {1986},
issn = {0021-9991},
doi = {https://doi.org/10.1016/0021-9991(86)90102-6},
url = {https://www.sciencedirect.com/science/article/pii/0021999186901026},
author = {Christopher R Anderson},}

@article{Zhang2014,
author = {Zhang, Xinxin and Bridson, Robert},
title = {A PPPM Fast Summation Method for Fluids and Beyond},
year = {2014},
issue_date = {November 2014},
publisher = {Association for Computing Machinery},
address = {New York, NY, USA},
volume = {33},
number = {6},
issn = {0730-0301},
url = {https://doi.org/10.1145/2661229.2661261},
doi = {10.1145/2661229.2661261},
journal = {ACM Trans. Graph.},
month = {nov},
articleno = {206},
numpages = {11}
}

@article{Li2016,
author = {Li, Dingzeyu and Fei, Yun and Zheng, Changxi},
title = {Interactive Acoustic Transfer Approximation for Modal Sound},
year = {2016},
issue_date = {December 2015},
publisher = {Association for Computing Machinery},
address = {New York, NY, USA},
volume = {35},
number = {1},
issn = {0730-0301},
url = {https://doi.org/10.1145/2820612},
doi = {10.1145/2820612},
journal = {ACM Trans. Graph.},
month = {dec},
articleno = {2},
numpages = {16}
}

@article{Mittal2005,
author = {Mittal, Rajat and Iaccarino, Gianluca},
title = {IMMERSED BOUNDARY METHODS},
journal = {Annual Review of Fluid Mechanics},
volume = {37},
number = {1},
pages = {239-261},
year = {2005},
doi = {10.1146/annurev.fluid.37.061903.175743},
URL = {https://doi.org/10.1146/annurev.fluid.37.061903.175743},
eprint = {https://doi.org/10.1146/annurev.fluid.37.061903.175743}
}

@article{MITTAL2008,
title = {A versatile sharp interface immersed boundary method for incompressible flows with complex boundaries},
journal = {Journal of Computational Physics},
volume = {227},
number = {10},
pages = {4825-4852},
year = {2008},
issn = {0021-9991},
doi = {https://doi.org/10.1016/j.jcp.2008.01.028},
url = {https://www.sciencedirect.com/science/article/pii/S0021999108000235},
author = {R. Mittal and H. Dong and M. Bozkurttas and F.M. Najjar and A. Vargas and A. {von Loebbecke}},
}

@article{bempp,
  title={Bempp-cl: A fast Python based just-in-time compiling boundary element library.},
  author={Betcke, Timo and Scroggs, Matthew W},
  journal={Journal of Open Source Software},
  volume={6},
  number={59},
  pages={2879},
  year={2021}
}

@article{Langlois2016,
author = {Langlois, Timothy R. and Zheng, Changxi and James, Doug L.},
title = {Toward Animating Water with Complex Acoustic Bubbles},
year = {2016},
issue_date = {July 2016},
publisher = {Association for Computing Machinery},
address = {New York, NY, USA},
volume = {35},
number = {4},
issn = {0730-0301},
url = {https://doi.org/10.1145/2897824.2925904},
doi = {10.1145/2897824.2925904},
journal = {ACM Trans. Graph.},
month = {jul},
articleno = {95},
numpages = {13}
}

@inproceedings{sigsoundcourse,
author = {James, Doug L.},
title = {Physically Based Sound for Computer Animation and Virtual Environments},
year = {2016},
isbn = {9781450342896},
publisher = {Association for Computing Machinery},
address = {New York, NY, USA},
booktitle = {ACM SIGGRAPH 2016 Courses},
articleno = {22},
numpages = {8},
location = {Anaheim, California},
series = {SIGGRAPH '16}
}

@article{shell,
author = {Chadwick, Jeffrey N. and An, Steven S. and James, Doug L.},
title = {Harmonic Shells: A Practical Nonlinear Sound Model for near-Rigid Thin Shells},
year = {2009},
issue_date = {December 2009},
publisher = {Association for Computing Machinery},
address = {New York, NY, USA},
volume = {28},
number = {5},
issn = {0730-0301},
journal = {ACM Trans. Graph.},
month = {dec},
pages = {1–10},
numpages = {10}
}

\end{document}